# ARTIFICIAL NEURAL NETWORK BASED POWER SYSTEM STABILIZER ON A SINGLE MACHINE INFINITE BUS SYSTEM MODELLED IN DIGSILENT POWERFACTORY AND MATLAB

Ali Kharrazi
School of Engineering and Information Technology
Murdoch University, Western Australia

## *ABSTRACT*

*In this paper the use of artificial neural network in power system stability is studied. A predictive controller based on two neural networks is designed and tested on a single machine infinite bus system which is used to replace conventional power system stabilizers. They have been used for decades in power system to dampen small amplitude low frequency oscillation in power systems. The increases in size and complexity of power systems have cast a shadow on efficiency of conventional method. New control strategies have been proposed in many researches. Artificial Neural Networks have been studied in many publications but lack of assurance of their functionality has hindered the practical usage of them in utilities. The proposed control structure is modelled using a novel data exchange established between MATLAB and DIgSILENT power factory. The result of simulation proves the efficiency of the proposed structure.*

## 1. Introduction

Small amplitude low frequency oscillations have been observed in power systems since 1960's [3]. They are the result of kinetic energy exchange between generation units. These oscillations have been known as electromagnetic mode of oscillation and if they do not dampen properly, they could affect the power quality or even cause the isolation of a part of power system. They may happen in two modes, local and inter area. Local mode is when one generator swings against the rest of the power system and inter area is when a group of generators in one area swing against a group of generator in another area. From control structure point of view, the phase lag exerted to the system as a result of the controller of the excitation system will originate these oscillations [12].

In order to dampen these oscillations a component in phase with speed deviation should be added to the mechanical torque. This has been achieved by use of conventional lead lag controllers known as Conventional Power System Stabilizer (CPSS). CPSS have been used in many utilities however their lack of flexibility to changes of power system has questioned their efficiency. Many new control strategies have been proposed to achieve more adaptive stabilizers.

Artificial intelligent techniques have been the focus of many researches [13]. In this paper a predictive controller based on two artificial neural networks is proposed. The proposed model is simulated on a novel data exchange platform using MATLAB and POWERFACOTRY data exchange.

Simulation software packages have been used by engineers in order to investigate the behaviour of systems and validate the functionality of new designs without running detrimental experiments on real systems. In Electrical powers engineering various useful software packages have been deployed by engineers, operators and students to study the behaviour of power systems in various operating conditions. Along with the advances in equipments and technologies in power systems, more complicated and enhanced simulation software tools with more capabilities and applications have been introduced. The interface between two simulations software where each software package is used for specific area makes the optimum use of them. The Dynamic Data Exchange supported by operating systems creates and environment where the real time data exchange between different software could be achieved [1].

In this paper a simulation platform using DIgSILENT PowerFactory and MATLAB Artificial Neural Network Toolbox is proposed. PowerFactory is a powerful simulation tool for power systems. Many conventional control loops could be modelled in PowerFactory but it does not support advanced and intelligent control strategies. On the other hand it is possible to link control blocks to MATLAB scripts with in control loop in PowerFactory [2]. In each time step during simulation a common data base between MATLAB and PowerFactory is shaped where PowerFactory exports desired data to MATLAB and calls the scripts which are saved as m file in MATLAB environment. When the codes have been executed PowerFactory imports the result of the calculations.

The power system under study is a Single Machine Infinite Bus (SMIB) modelled in PowerFactory which is linked to Artificial Neural Network based Power System Stabilizer coded in MATLAB using Neural Network Toolbox.

In next section of the report the application of Power System stabilizer (PSS) and different control strategies used as PSS will be introduced. Next section describes the Power system model in PowerFactory. In section 4 the data interface between the two software packages is explained. Section 5 explains the Artificial Neural Network based Power System Stabilizer (ANNPSS) created in MATLAB. Next section presents some simulation and results followed by conclusion.

## 2. Power system stabilizer

Power system stabilizers have been used for decades in power systems to enhance the transient stability of power system [3]. They are employed in order to dampen the small amplitude low frequency oscillations observed in power systems. These oscillations in rotor speed and rotor angle of generation units are result of phase lag imposed by excitation control system, generation and power system.. The frequency of these oscillations is in range of 0.5 to 3 Hz. They may persist in the power system and if does not dampen properly they may result in poor power quality or even the isolation of generation unit or a part of power system.

In order to add enough dampening to the system, a component in phase with speed deviation is added to electrical torque in rotating mass. The conventional power system stabilizers are embedded through Automatic Voltage Regulator and the input is speed or speed deviation of the rotor. The output is stabilizer signal (UPSS). In Figure1 the block diagram of a conventional power system stabilizer (CPSS) is shown.

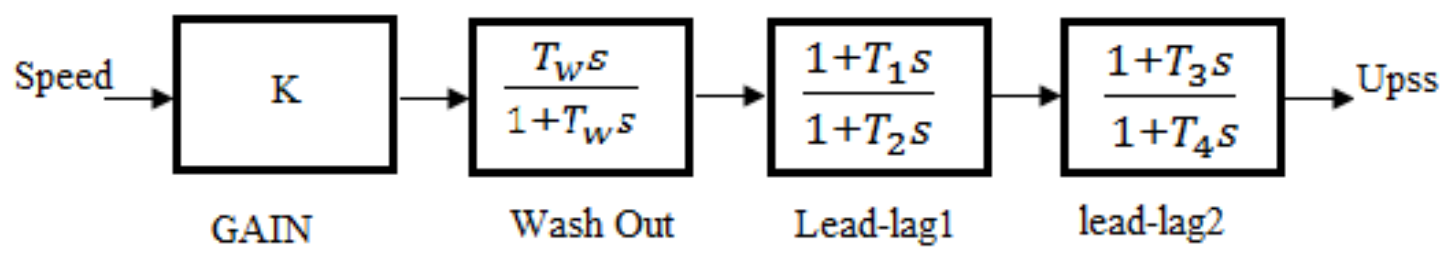

Figure 1 Conventional Power System Stabilizer

As seen in Fig.1 the CPSS consists of a gain block, a washout filter to filter out the steady state voltage and two lead lag blocks to compensate the phase lag of the system. The parameters of lead lags block should be tuned properly to achieve desired responses.

Although CPSS has been used widely in many utilities with good outcomes, they suffer from drawbacks of using conventional controllers such as dealing with uncertainties. On the other the hand complexity of power systems and its nonlinear nature as well as constantly changing parameters of the system has motivated researches to propose more advance control strategies.

Many adaptive and robust control structures have been proposed to be used as Power System Stabilizer [4] [5].  In many publications Artificial Intelligent control techniques have been used to enhance the stability of power systems including tuning of the parameters of CPSS [6] or replacing the conventional controller with Artificial Neural Network (ANN) structures  [7]  [8]. The flexibility and adaptablity of ANNs due to large number of parametrs and training capability, have propmted researches to introduced Artificial Neural Network Based Power System Stabilizers.

## 3. Power System under Study

The power system in this paper is a Single Machine Infinite Bus which is modelled in PowerFactory shown in Figure 2. This topology is used to study the local mode of oscillation. It is a single generator unit connected to an infinite bus through a transmission line. The infinite bus is the equivalent of the rest of power system in form of a bus with fix voltage.

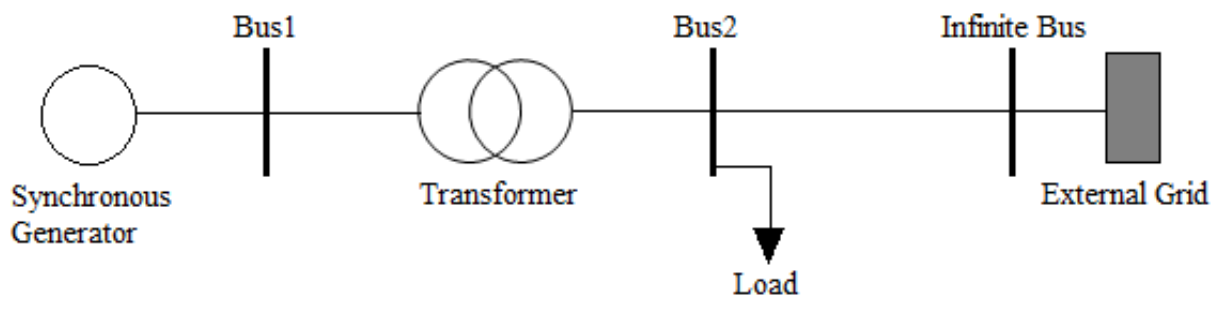

Figure 2 Single Machine Infinite

The control block of synchronous generator is modelled as composite model in PowerFactory. Different control structure such as governor, AVR and PSS are embedded in composite model. This structure is shown in Figure 3.

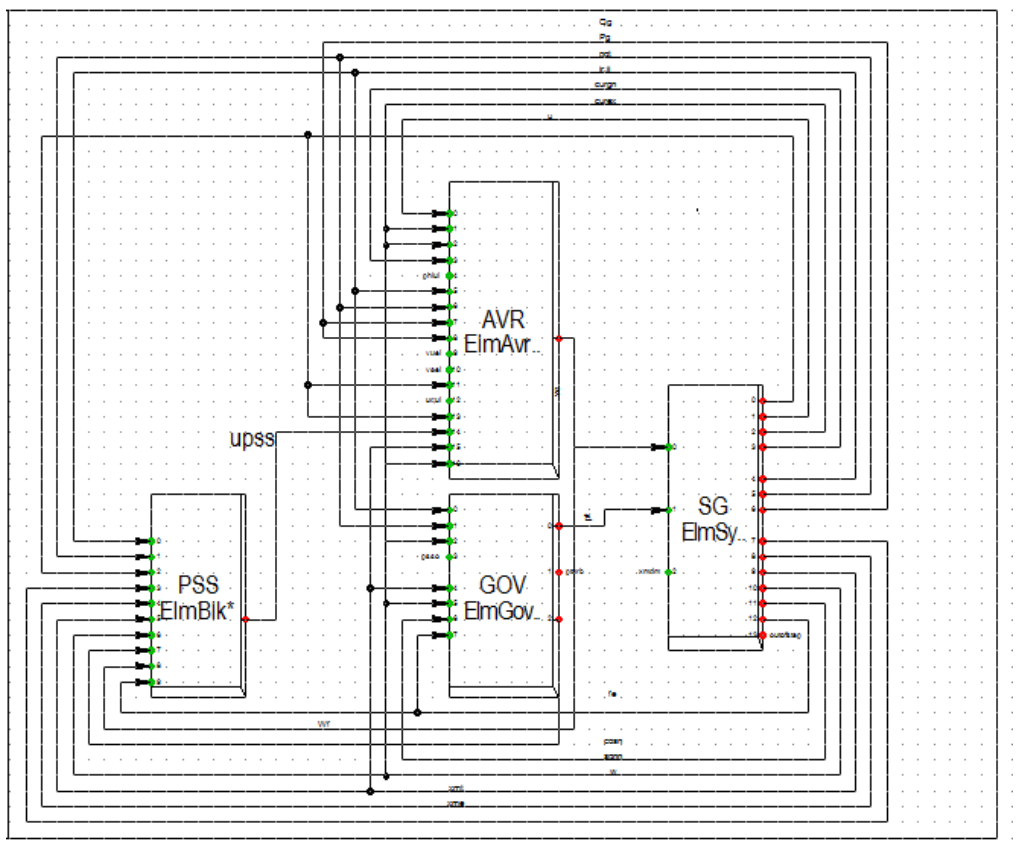

Figure 3 Composite model Signal Interconnection

Each slot in the composite model is linked to a control block known as common block. Now the PSS block is linked to a CPSS shown in Fig. 1. This composite model could be changed and modified by user. New blocks could be added and the signal interconnection could be changed.

## 4. PowerFactory MATLAB Interface

Classic control structure could be modelled in PowerFactory. These models could be embedded in composite model shown in Fig. 3. But more complicated control structure could not be simulated in PowerFactory. These constrains in simulation software have prompted the interface of different simulation tools where each tool has a specific application. In [1] the process and requirments of interfacing transinet simulation software which use mathematical algorithm to simulate electromagnetic phenomena are illustrated.

In this paper a Neural Network based controller is replaced with the CPSS in Figure 3. A common block is added to the composite model shown in Figure 3, which is linked to a MATLAB script. In [2] the process to link a common model in PowerFactory to MATLAB script is illustrated. The new signal interconnection with MATLAB interface block is shown in Figure 4.

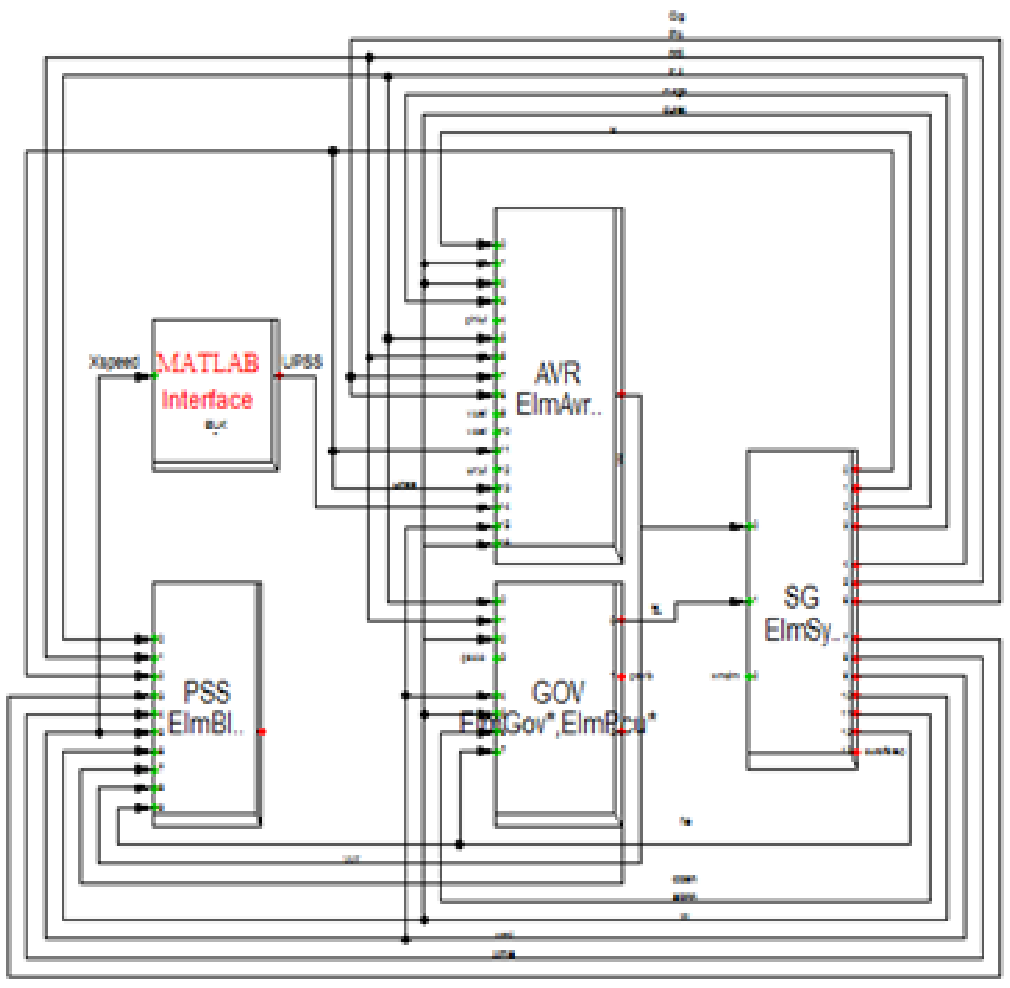

Figure 4 New Block for MATLAB Interface

Since MATLAB codes are to replace PSS in control loop, signals connected to this new block are speed from Synchronous Generator and UPSS to the AVR block. In order to test the proper functionality of the data interface the block diagram shown in Figure 1 was modelled in MATLAB Simulink and the result of simulation were compared with the block modelled in PowerFactory. The block modelled in Simulink is shown in figure 5.

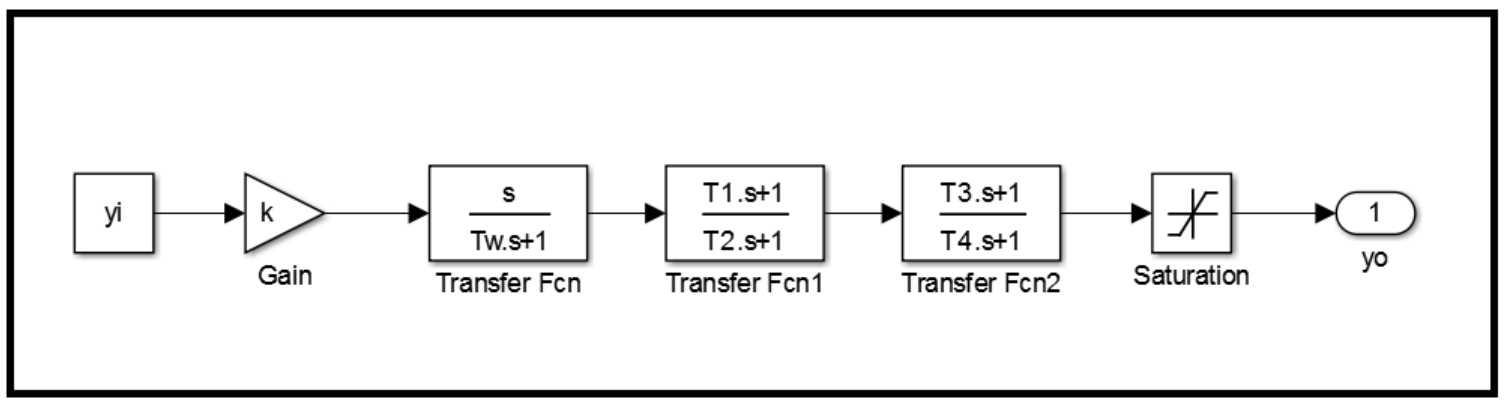

Figure 5 CPSS modelled in SIMULINK

The validation of correct data exchange is done through a simulation of three phase fault in the transmission line. The result of two CPSS block from MATLAB and POWEFACTORY are shown in figure 6.

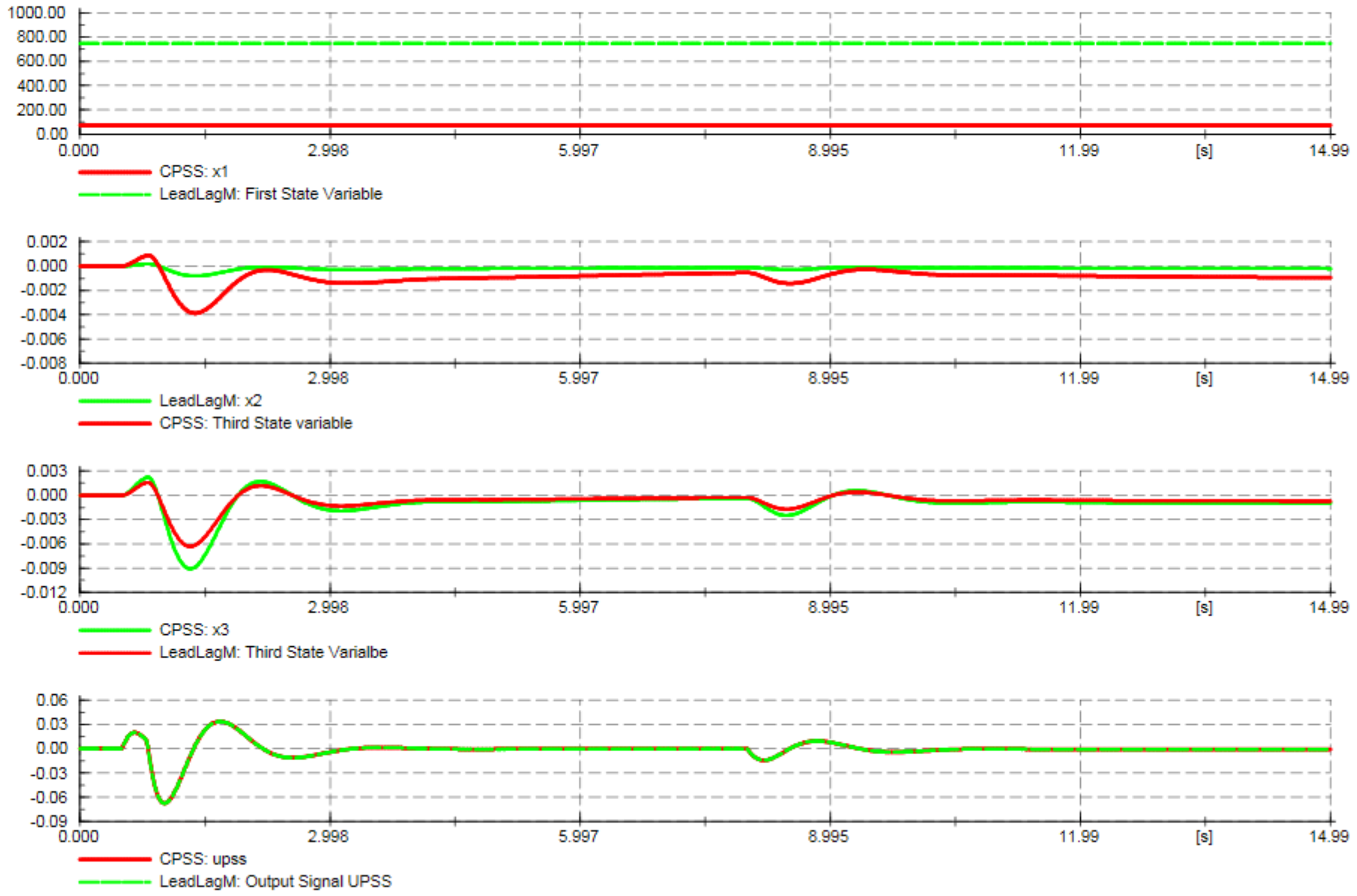

Figure 6 Result of Simulation Comparing Simulink and PowerFactory Blocks

It could be seen in Figure 6 that the simulation result done through data exchange between POWERFACTORY and yield identical results to the simulation done by blocks in POWERFACTORY.

## 5. Artificial Neural Network based Power System Stabilizer (ANNPSS)

Although conventional and classic control structure could improve the transient stability of power systems, but lack of adaptivity and inherent drawbacks in the design such as uncertainty has cast a shadow on their performance. In addition the increasing complexity and demand of power system has pushed them to operate at stability limits. Thus new intelligent and adaptive controllers are inevitable.

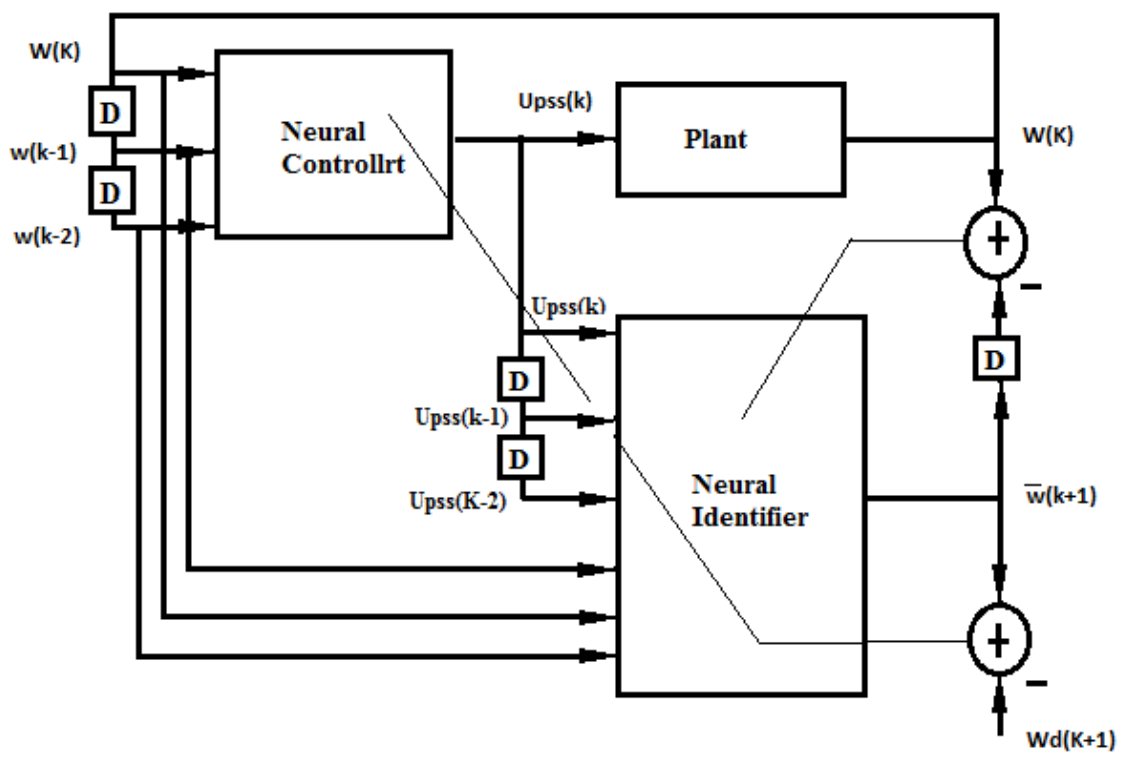

Figure 7 ANNPSS Control Structure

Controllers based in ANN have been proposed in many publications. In this paper the control strategy proposed in [9] is adapted to enhance the stability of power system. The proposed structure is a predictive controller which use two neural networks in order to achieve control goal in nonlinear systems: Neural Controller and Neural Identifier. The proposed structure is to design an optimal controller to minimise a cost fucntion. Same topology has been used in [10] to realize a power system stabilzier. The novelty in this paper is the software tools employed which eases the training and simulation of the proposed system. The control structure of ANNPSS is shown in Figure 5.

The Plant in Figure 7 is the power system as seen from point of view of PSS. The input to the plant is the stabilizer signal and the output is the speed. This control structure is modelled in MATLAB and is linked to PowerFactory. In composite model in PowerFactory two interfaces are added for each Neural Network as shown in Figure 8. The two neural networks in the control loop are illustrated in the following sections.

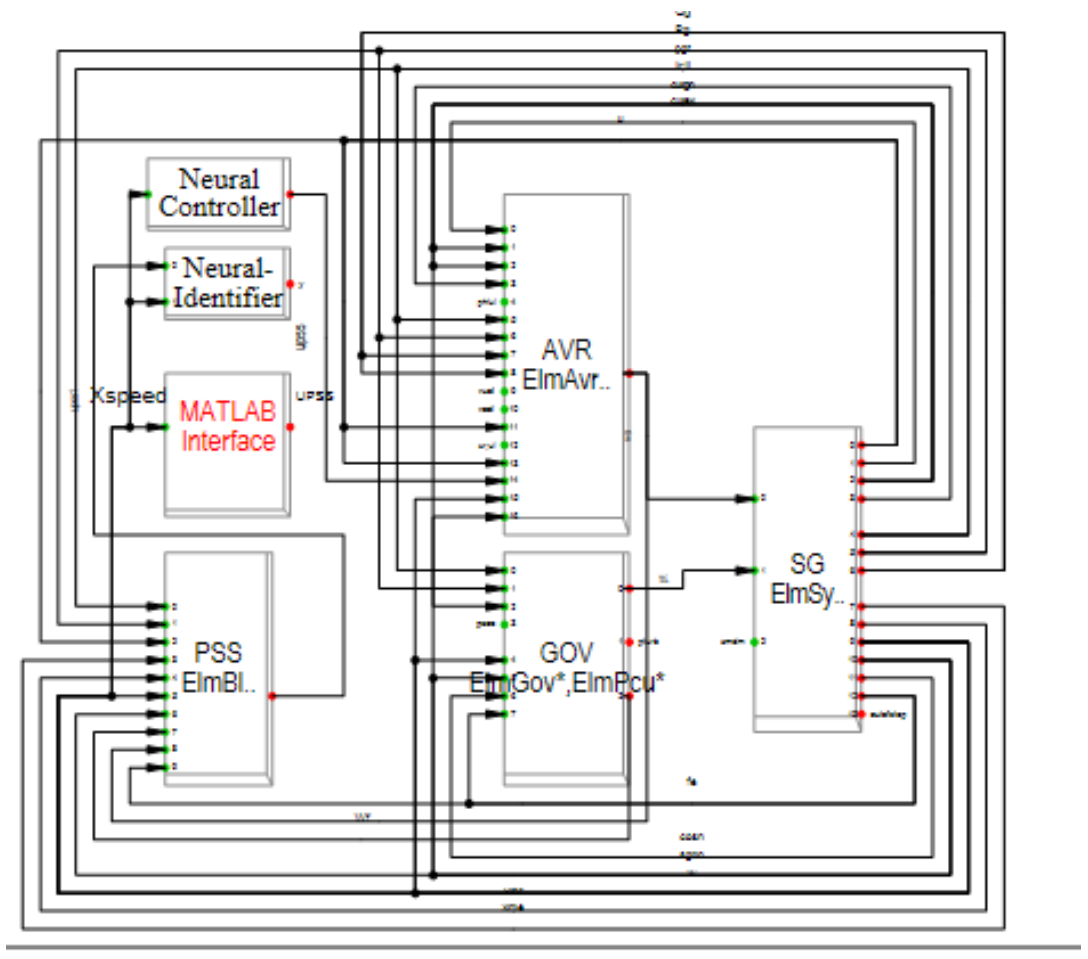

Figure 8 Neural Controller and Identifier Interface between MAT|LAB and PowerFactory

## 5.1 Neural Identifier

The main aim of Neural Identifier (NI) is to approximate the behaviour of the system. In order to achieve this goal Non-linear Auto-Regressive Moving Average (NARMA) [11] model which is a representation of nonlinear sytems is used. Based on NARMA model the output of a discrete system is a function of the current and previous values of system inputs and outputs as shown in Equation 1.

$$\hat{y}(k+1) = f\left(\begin{bmatrix} y(k), y(k-1) \dots, y(k-n) \\ u(k), u(k-1),, \dots u(k-n) \end{bmatrix}\right) \tag{1}$$

The NARMA model could be realized by feed-forward Neural Network which is modelled using MATLAB Neural Network Toolbox (nntool). A Nonlinear Autoregressive with External Input network (narxnet) including one hidden layer with ten neurons and output layer with one neuron is created in MATLAB. This network could be seen in Figure11.

The task of this network is to learn to predict the output of the plant using the current and previous values of input and output which are stabilizer signal and speed of rotor. These signals are fed to the Neural Identifier Block in composite model shown in Figure6. The two following sections illustrate the training and validating the identifier.

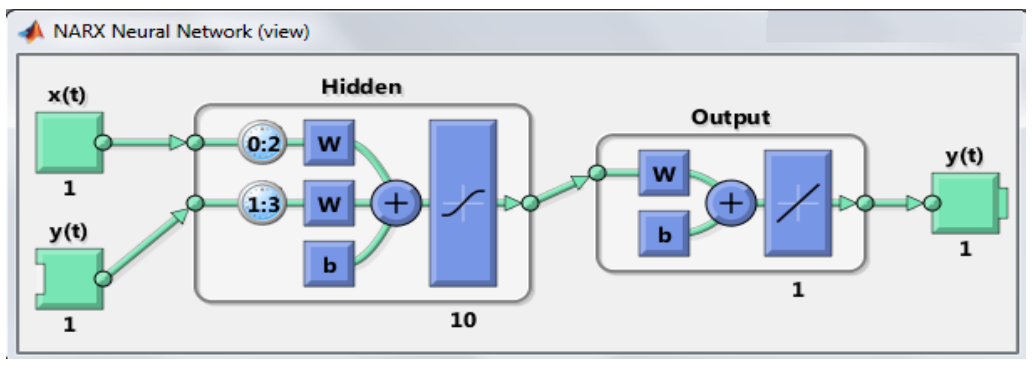

Figure 9 Neural Identifier Network created by nntool

### 5.1.1 Training Neural Identifier

The NI should be trained with adequate learning data in order to be able to predict the output of the plant. The learning data which include stabilizer signal and speed should be chosen adequately to give enough information for proper learning. The signals used, should excite all oscillatory modes of the system.   In order to achieve this MATLAB's *idinput* command which generates signal for identification is used. The signal shown in Figure 12 is created using the mentioned command. This signal is sum of sinusoid signals with different frequencies in order to excite all the modes of the plant.

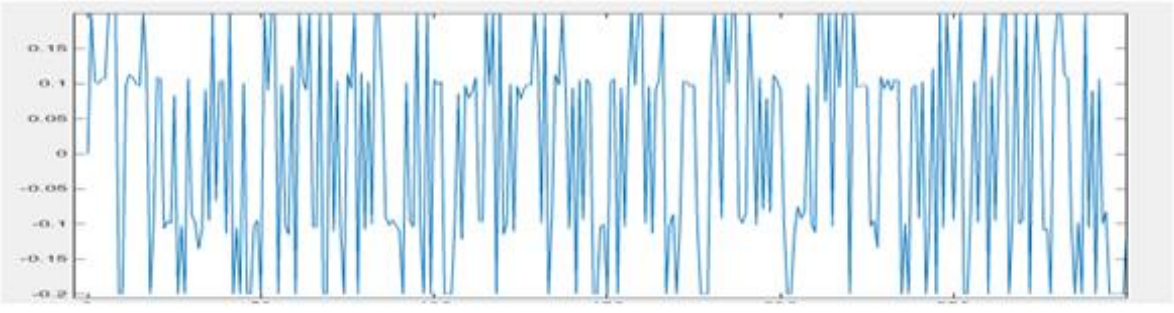

Figure 10 Signal Generated for Identifier Training

This signal is send to PowerFactory using the MATLAB Interface block shown in Figure 10 as UPSS signal during simulation. After running the simulation the speed of generator signal is saved as CSV file in PowerFactory. This CSV file is loaded in MATLAB and is used as the target data for offline training of Neural Identifier along with the signal shown in Figure 10. The training process is done using *train* command in MATLAB. Online training could be performed as well. In order to perform online training the Neural Identifier network should be embedded into the MATLAB codes linked to Neural Identifier Block in Figure 8. The input signals to the Neural Identifier Blocks are the signal coming from MATLAB Interface Block and speed signal coming from Synchronous Generator Block.

The cost function of training is shown in Equation 2.

$$J(k) = \frac{1}{2}(w(k) - \hat{w}(k))^2 \qquad (2)$$

In Equation 2 w(k) is the output signal (rotor speed) of the real plant in modelled in PowerFactory and $\hat{w}(k)$ is the output of Neural Identifier.

### 5.1.2 Validating Neural Identifier

In order to validate proper training of Neural Identifier, a simulation is run to compare the output of the NI in MATLAB and the plant in PowerFactory. In the simulation a three phase fault in transmission line in 0.2s with duration of 500 ms and a 10 % for increase in load in 8s are scheduled.  The result of simulation is shown in Figure 11. As seen in Figure 11 the output of Neural Identifier is similar to the speed signal in PowerFactory.

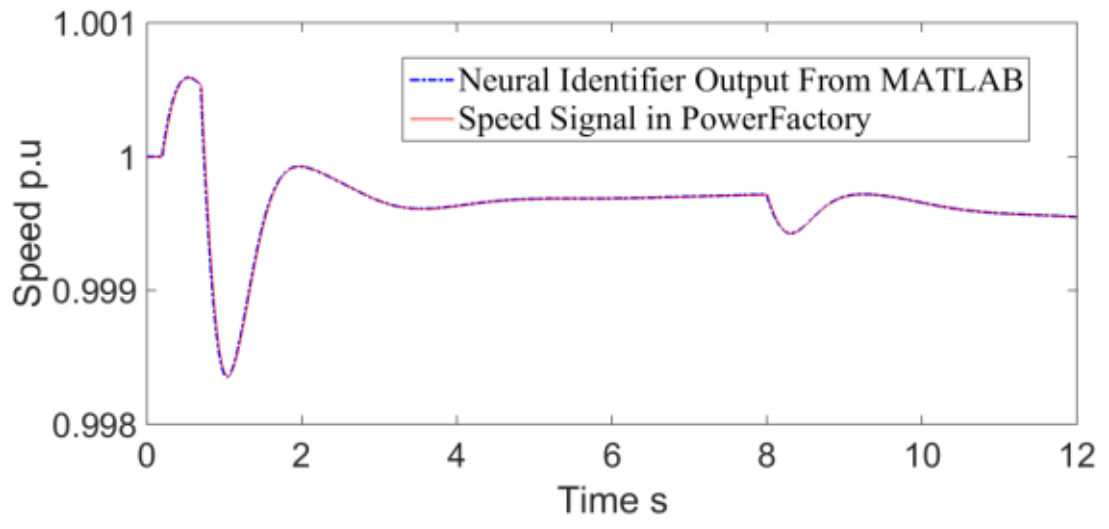

Figure 11 Comparison of Neural Identifier Output and PowerFactory Model

## 5.2 Neural Controller

The Neural Controller (NC) should be trained to create the optimum stabilizer signal. A time delay Neural Network created using nntool in MATLAB, is used as Neural Controller shown in Figure 12.

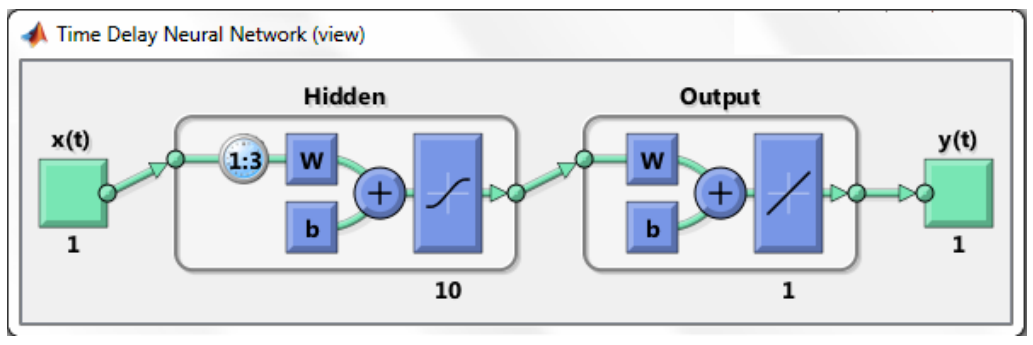

Figure 12 Neural Controller

This network is embedded in MATLAB codes which are linked to the Neural Controller Block in Figure 8. The training of NC is done using the NI output. The NI could predict one step ahead of future value of the system response using the previous stabilizer signal and current and previous speed signal. The predicted value is compared with desired value which in this case is 1 p.u since we want the speed to be fix at all times. So the cost function for training of NC could be defined as shown in Equation 3.

$$J(k) = \frac{1}{2}\left(W_d(k+1) - \widehat{w}(k+1)\right)^2 \qquad (3)$$

In Equation 3 the cost function is the mean square error of desired value compared to NI prediction. Since the cost function in Eq.3 is not a function of NC output, the training algorithm in MATLAB nntool could not be used which use the derivative of cost function w.r.t. network output. In order to train the NC the output mean square error of NI is defined and this error is propagated back through NI using chain rule. The output error and back propagated error are shown in figure 13.

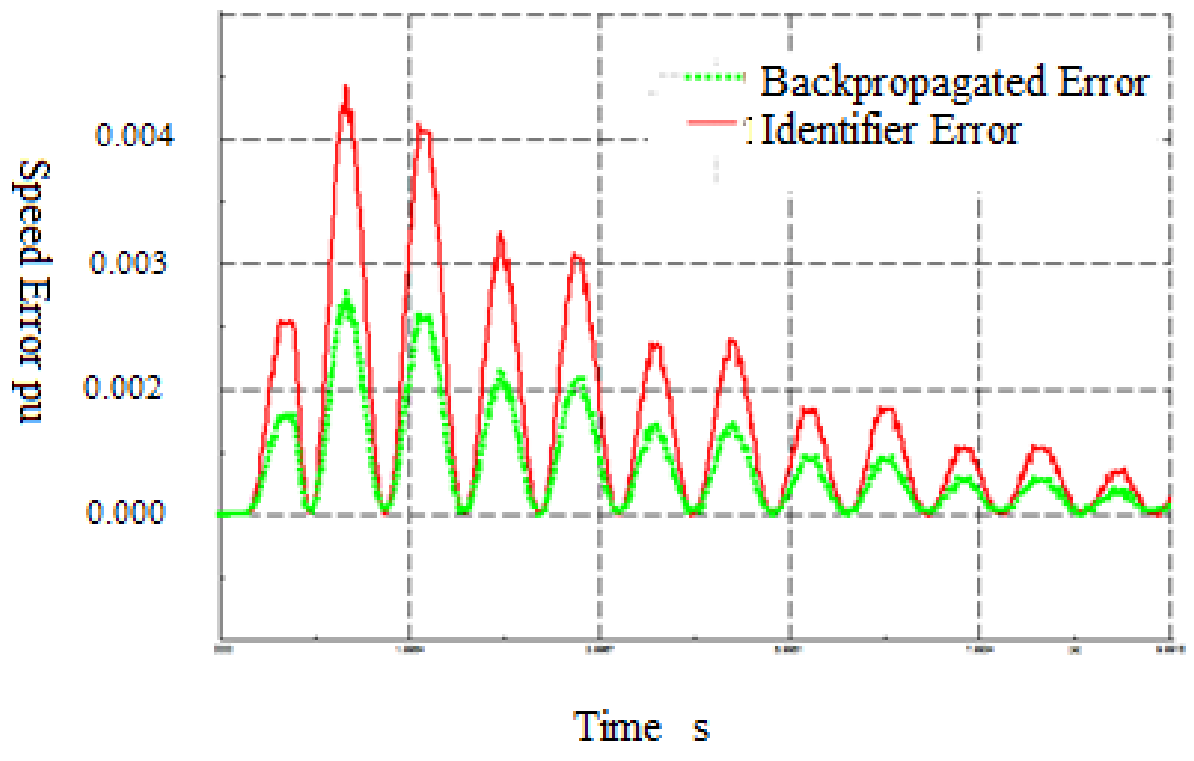

Figure 13 Error of Identifier Output with Desired Value and Back Propagated Error

This process is embedded in MATLAB codes linked to Neural Identifier Block. In each time step the signal is fed to the system as UPSS signal. NI predicts the next time step of output of the system and the mean square error of predicted value and desired value is calculated. This error is back propagated through NI and the error of stabilizer signal which is the output of NC is obtained. This Signal is used to train the NC.

# 6. Simulation and Results

In order to validate and test the functionality of the proposed control structure some simulations are performed. After training of NC the Neural Identifier Block is disabled and the output of Neural Controller block which is linked to MATLAB is connected to AVR block as UPSS signal. All the simulations are run in PowerFactory with active data interface with MATLAB. In the first simulation shown in Figure 14, a three phase fault lasting for 500 ms at 0.2s and 10 % increase in load in 8s are scheduled.

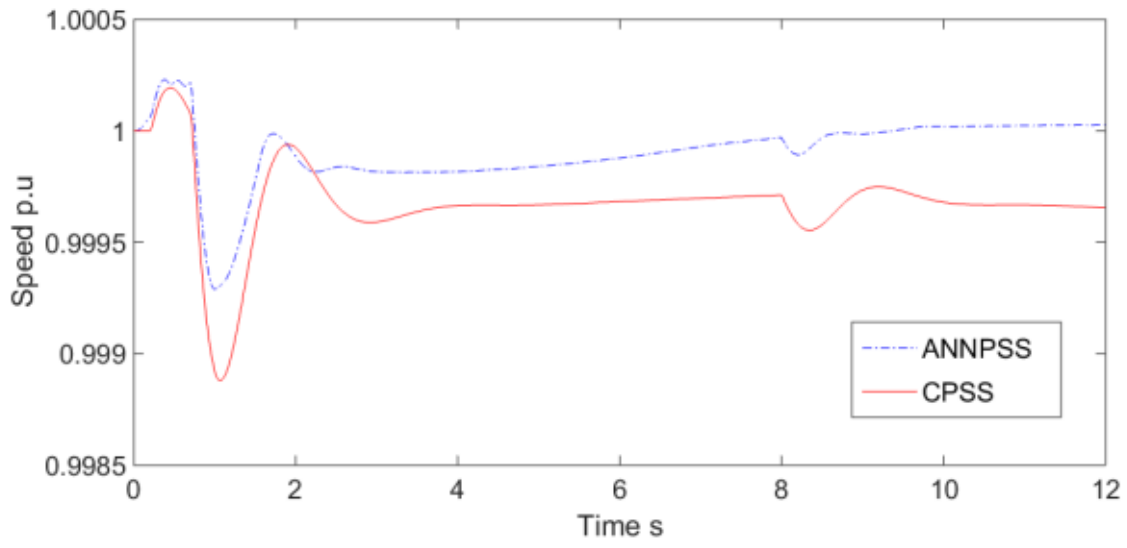

Figure 14 ANNPSS and CPSS Comparison to 500 ms Three Phase Fault and 10 % load increase

In Figure 15 12 the result of simulation of three phase fault which lasts for 1s is shown.

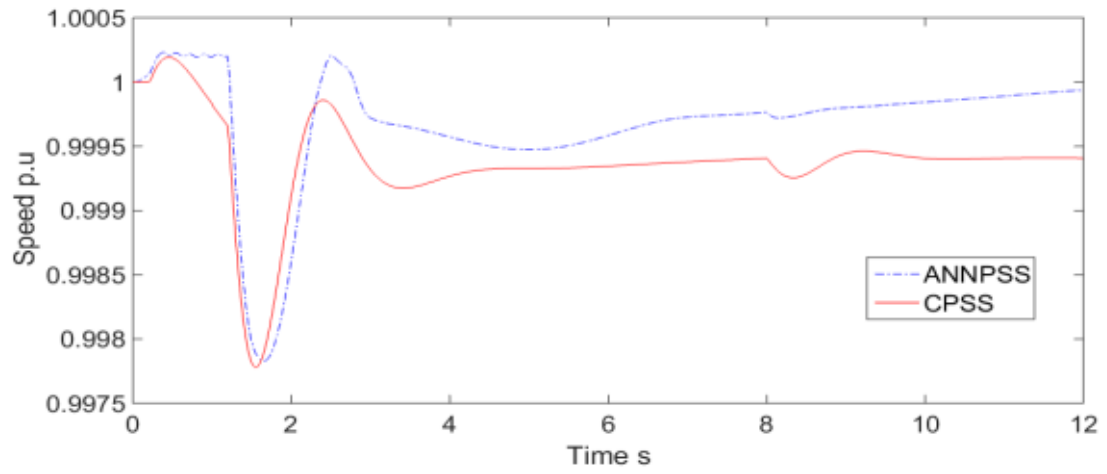

Figure 15 ANNPSS and CPSS Comparison to 1s Three Phase Fault

The last simulation is performed while the local load of the power system is in maximum. The results are shown in Figure 16.

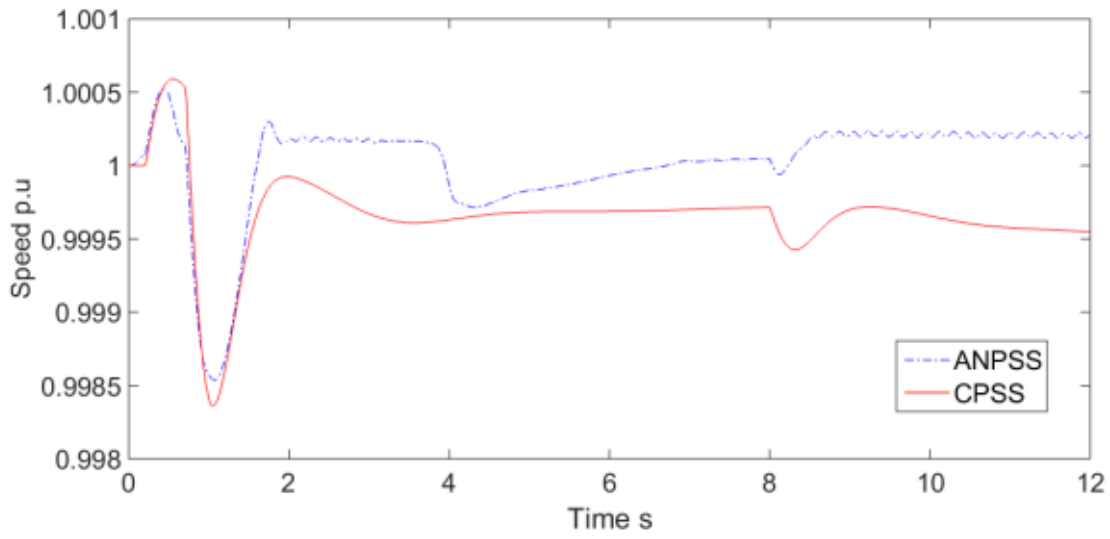

Figure 16 ANNPSS and CPSS Comparison to 500 ms Three Phase Fault in Full Loading

In all the simulation the response of rotor speed to CPSS modelled in PowerFactory is compared with response to ANNPSS modelled in MATLAB. It could be seen from the figures that ANNPSS has better response and could regain steady state in a shorter time.

## 7. Conclusion

In this paper a new platform for data interface between MATLAB and DIgSILENT PowerFactory was employed to run the simulation of an Artificial Neural Networked based Power System Stabilizer. The interface between the two software packages makes optimum use of each one in their specific area of applications. The proposed Neural Network control strategy is modelled and trained in MATLAB Environment using the data shared with PowerFactory during simulation. The result of simulation validates the effectiveness of proposed control structure while the data interface facilitates the process of simulation.